\documentstyle[12pt,axodraw]{article}

\newcommand{\re}{\mbox{$\rm e$}}
\newcommand{\ri}{\mbox{$\rm i$}}

\newcommand{\bfm}[1]{\mbox{\boldmath$#1$}}
\newcommand{\ratio}[2]{\mbox{$#1\over#2$}}
\newcommand{\tr}{\mbox{$\,\rm Tr$}}

\newcommand{\ra}{\rightarrow}

\begin{document}
\baselineskip=17pt
\parskip=5pt
  
\begin{titlepage} \footskip=5in     

\title{ 
\begin{flushright} \normalsize 
ISU-HET-98-5 \vspace{0.1em} \\ DOE/ER/40561-35-INT-98 \vspace{1em} \\
November 1998 \vspace{5em} \\  
\end{flushright}
\large\bf   
$\bfm{C\!P}$ Violation in Non-Leptonic $\Omega^-$ Decays
}  

\author{\normalsize\bf 
Jusak~Tandean\thanks{E-mail: jtandean@iastate.edu} \ and  
{G.~Valencia\thanks{E-mail: valencia@iastate.edu}}  \\
\normalsize\it Department of Physics and Astronomy, Iowa State University, 
Ames, IA 50011 \\
\normalsize\it and \\   
\normalsize\it Institute for Nuclear Theory, University of Washington, Box 
351550, \\
\normalsize\it Seattle, WA 98195     
}

\date{}
\maketitle

\begin{abstract}
We estimate the size of the $C\!P$-violating rate asymmetry for the 
decay  $\Omega^-\rightarrow\Xi\pi$. Within the standard model 
we find a value of $\,2\times 10^{-5}$, and it could be as much as 
ten times larger if new physics is responsible for $C\!P$ violation. 
Even though our calculation suffers from the usual uncertainty in 
the estimate of hadronic matrix elements, we find a rate asymmetry 
that is significantly larger than the corresponding one for 
octet-hyperon decays.
\end{abstract}

\end{titlepage}

\section{Introduction}

In this paper we estimate the size of the $C\!P$-violating rate 
asymmetry for $\,\Omega^- \rightarrow \Xi \pi\,$ decays. As is 
well known, such rate asymmetries for octet-hyperon decays 
are small as a result of the 
product of three small factors: a ratio of  $\,|\Delta\bfm{I}|=3/2\,$  
to  $\,|\Delta\bfm{I}|=1/2\,$  amplitudes;  
a small strong-rescattering phase;  
and a small $C\!P$-violating phase~\cite{donpa}.  
We find that for the decay channel  $\,\Omega^-\rightarrow\Xi\pi\,$  
all of these factors are larger than their counterparts for 
octet-hyperon decays, and this results in a rate asymmetry 
that could be as large as  $\,2 \times 10^{-5}\,$  within the minimal 
standard model.  
Physics beyond the standard model could enhance this rate asymmetry 
by a factor of up to ten.  
Our calculation suffers from typical hadronic uncertainties in 
the computation of matrix elements of four-quark operators and 
for this reason it should be regarded as an order-of-magnitude 
estimate.

\section{$\Omega^- \rightarrow \Xi\pi$ decay}
  
The measured decay distributions of these decays are consistent 
with the amplitudes being mostly P-wave~\cite{pdb}.  
We parametrize the P-wave amplitude in the form
\begin{eqnarray}   \label{amplitude}     
\ri {\cal M}_{\Omega^-\rightarrow\Xi\pi}^{}  \;=\;  
G_{\rm F}^{} m_{\pi}^2\; \bar{u}_\Xi^{}\, 
{\cal A}_{\Omega^-\Xi\pi}^{\rm (P)}\, k_\mu^{}\, u_\Omega^\mu   
\;\equiv\;  
G_{\rm F}^{} m_{\pi}^2\; 
{\alpha_{\Omega^-\Xi}^{\rm (P)}\over \sqrt{2}\, f_{\!\pi}^{}}\,  
\bar{u}_\Xi^{}\, k_\mu^{}\, u_\Omega^\mu      \;,
\end{eqnarray}    
where  the  $u$'s  are baryon spinors,  $k$  is the outgoing 
four-momentum of the pion, and  $f_{\!\pi}^{}$  is the pion-decay 
constant.  
The P-wave amplitude has both  $\,|\Delta\bfm{I}|=1/2\,$  and 
$\,|\Delta\bfm{I}|=3/2\,$  components which are, in general, complex. 
We write
\begin{eqnarray}   
\begin{array}{c}   \displaystyle   
\alpha_{\Omega^-\Xi^0}^{\rm (P)}  \;=\;   
\ratio{1}{\sqrt{3}} \left( 
\sqrt{2}\, \alpha^{(\Omega)}_1  
\re^{{\rm i}\delta_1^{} + {\rm i}\phi_1^{}}
\,-\,  \alpha^{(\Omega)}_3 \re^{{\rm i}\delta_3^{} + {\rm i}\phi_3^{}}
\right)   \;,  
\vspace{2ex} \\   \displaystyle
\alpha_{\Omega^-\Xi^-}^{\rm (P)}  \;=\;   
\ratio{1}{\sqrt{3}} \left( 
\alpha^{(\Omega)}_1 \re^{{\rm i}\delta_1^{} + {\rm i}\phi_1^{}} 
\,+\, \sqrt{2}\, \alpha^{(\Omega)}_3  
\re^{{\rm i}\delta_3^{} + {\rm i}\phi_3^{}}
\right)   \;,   
\end{array}   \label{isolabels}
\end{eqnarray}  
where  $\alpha^{(\Omega)}_{1,3}$  are real quantities,  
strong-rescattering phases of the  $\Xi\pi$  system with  $\,J=3/2$,  
P-wave and  $\,I=1/2, 3/2\,$  quantum numbers are denoted by  
$\delta_{1}$, $\delta_{3}$,  respectively, and    
$C\!P$-violating weak phases are labeled  $\phi_{1}$,  $\phi_{3}$. 
The corresponding expressions for the antiparticle decay 
$\,\overline{\Omega}{}^-\rightarrow \overline{\Xi}\pi\,$  are 
obtained by changing the sign of the weak phases $\phi_{1}$, $\phi_{3}$ in 
Eq.~(\ref{isolabels}).

Summing over the spin of the  $\Xi$  and  averaging over the spin of 
the  $\Omega^-$,  one derives from  Eq.~(\ref{amplitude})  
the decay width  
\begin{eqnarray}   \label{width''}     
\Gamma(\Omega^-\ra\Xi\pi)  \;=\;  
{|\bfm{k}|^3 m_{\Xi}^{}\over 6\pi m_{\Omega}^{}} 
\Bigl| {\cal A}_{\Omega^-\Xi\pi}^{\rm (P)} \Bigr|^2 \, 
G_{\rm F}^2 m_{\pi}^4   \;.  
\end{eqnarray}    
As was found in  Ref.~\cite{jusak}, using the measured decay 
rates~\cite{pdb} and  ignoring all the phases, we can extract the ratio  
$\, \alpha_{3}^{(\Omega)}/\alpha_{1}^{(\Omega)} =-0.07\pm 0.01.\,$  
Final-state interactions enhance this value, but this enhancement 
is not significant for the values of the scattering phases that we 
estimate in the following section. 
This ratio is higher than the corresponding ratios in other hyperon 
decays~\cite{atv},  which range from  $0.03$  to  $0.06$ in magnitude, 
and provides an enhancement factor for the $C\!P$-violating rate 
asymmetry in this mode.

By comparing the hyperon and anti-hyperon decays, we can construct 
$C\!P$-odd observables.  
The one considered here is the rate asymmetry
\begin{eqnarray}
\Delta \bigl( \Xi^0\pi^- \bigr)  &\!\!\equiv&\!\!  
{ \Gamma \bigl( \Omega^-\ra\Xi^0\pi^- \bigr) - 
 \Gamma \bigl( \overline{\Omega}{}^-\ra\overline{\Xi}{}^0\pi^+ \bigr)  
 \over  
 \Gamma \bigl( \Omega^-\ra\Xi^0\pi^- \bigr) + 
 \Gamma \bigl( \overline{\Omega}{}^-\ra\overline{\Xi}{}^0\pi^+ \bigr) }  
\nonumber \\  
&\!\!\approx&\!\!  
\sqrt{2}\; {\alpha_{3}^{(\Omega)}\over \alpha_{1}^{(\Omega)}}\; 
\sin \bigl( \delta_3^{}-\delta_1^{} \bigr) \, 
\sin \bigl( \phi_3^{}-\phi_1^{}\bigr)   \;,  
\label{cpobs}  
\end{eqnarray}  
where in the second line we have kept only the leading term in 
$\,\alpha_{3}^{(\Omega)}/\alpha_{1}^{(\Omega)}.\,$   
Similarly, 
$\, \Delta \bigl( \Xi^-\pi^0 \bigr) 
= -2 \Delta \bigl( \Xi^0\pi^- \bigr) .\,$ The current 
experimental results indicate that any D-waves are very small in 
these decays, and that the parameter $\alpha$ that describes 
P-wave--D-wave interference is consistent with zero: 
$\,\alpha \bigl( \Xi^0\pi^- \bigr) =0.09\pm 0.14\,$ and 
$\,\alpha \bigl( \Xi^-\pi^0 \bigr) =0.05\pm 0.21\,$~\cite{pdb}. For this 
reason we do not discuss the potential $C\!P$-odd asymmetry in 
this parameter.

\section{$\Xi\pi$-scattering phases}

There exists no experimental information on the $\Xi\pi$-scattering  
phases, and so we will estimate them at leading order 
in heavy-baryon chiral perturbation theory. 
The leading-order chiral Lagrangian for the strong interactions of 
the octet and decuplet baryons with the pseudoscalar octet-mesons 
is~\cite{manjen}
\begin{eqnarray}   \label{L1strong}
{\cal L}^{\rm s}  &\!\!=&\!\!      
\ratio{1}{4} f^2\,
 \tr \!\left( \partial^\mu \Sigma^\dagger\, \partial_\mu \Sigma \right) 
\;+\;  
\tr \!\left( \bar{B}_v^{}\, \ri v\cdot {\cal D} B_v^{} \right)    
\nonumber \\ && \!\!\! \!   
+\; 
2 D\, \tr \!\left( \bar{B}_v^{}\, S_v^\mu 
 \left\{ {\cal A}_\mu^{}\,,\, B_v^{} \right\} \right)     
+ 2 F\, \tr \!\left( \bar{B}_v^{}\, S_v^\mu\,    
 \left[ {\cal A}_\mu^{}\,,\, B_v^{} \right] \right)   
\nonumber \\ && \!\!\! \!   
-\;    
\bar{T}_v^\mu\, \ri v\cdot {\cal D} T_{v\mu}^{}  
+ \Delta m\, \bar{T}_v^\mu T_{v\mu}^{}  
+ {\cal C} \left( \bar{T}_v^\mu {\cal A}_\mu^{} B_v^{} 
                   + \bar{B}_v^{} {\cal A}_\mu^{} T_v^\mu \right)    
+ 2{\cal H}\; \bar{T}_v^\mu\, S_v^{}\cdot{\cal A}\, T_{v\mu}^{}   \;,       
\end{eqnarray}      
where we follow the notation of Ref.~\cite{manjen}.

\begin{figure}[t]         
   \hspace*{\fill} 
\begin{picture}(140,50)(-70,-20)    
\ArrowLine(-40,0)(0,0) \DashLine(-20,20)(0,0){3} 
\DashLine(0,0)(20,20){3} \ArrowLine(0,0)(40,0) \Vertex(0,0){3} 
\end{picture}
   \hspace*{\fill} 
\begin{picture}(140,50)(-70,-20)    
\ArrowLine(-40,0)(-20,0) \DashLine(-20,20)(-20,0){3} 
\ArrowLine(-20,0)(20,0) \DashLine(20,0)(20,20){3}   
\ArrowLine(20,0)(40,0) \Vertex(-20,0){3} \Vertex(20,0){3}  
\end{picture}
   \hspace*{\fill} 
\begin{picture}(140,50)(-70,-20)    
\ArrowLine(-40,0)(-20,0) \Line(-20,1)(20,1) \Line(-20,-1)(20,-1) 
\ArrowLine(-1,0)(1,0) \DashLine(-20,20)(-20,0){3} 
\DashLine(20,0)(20,20){3} \ArrowLine(20,0)(40,0) 
\Vertex(-20,0){3} \Vertex(20,0){3}  
\end{picture}  
   \hspace*{\fill} 
\\ 
   \hspace*{\fill} 
\begin{picture}(10,10)(-5,-5)
\Text(0,5)[c]{(a)}   
\end{picture}
   \hspace*{\fill} 
\\       
   \hspace*{\fill} 
\begin{picture}(140,50)(-70,-20)    
\ArrowLine(-40,0)(-20,0) \DashLine(-20,20)(-20,0){3} 
\ArrowLine(-20,0)(20,0) \DashLine(20,0)(20,20){3}   
\ArrowLine(20,0)(40,0) \Vertex(-20,0){3} \Vertex(20,0){3}  
\end{picture}
   \hspace*{\fill} 
\begin{picture}(140,50)(-70,-20)    
\ArrowLine(-40,0)(-20,0) \DashLine(-20,20)(0,10){3} 
\DashLine(0,10)(20,0){3} \ArrowLine(-20,0)(20,0) 
\DashLine(-20,0)(0,10){3}  \DashLine(0,10)(20,20){3}   
\ArrowLine(20,0)(40,0) \Vertex(-20,0){3} \Vertex(20,0){3}  
\end{picture}
   \hspace*{\fill} 
\\       
   \hspace*{\fill} 
\begin{picture}(140,50)(-70,-20)    
\ArrowLine(-40,0)(-20,0) \Line(-20,1)(20,1) 
\Line(-20,-1)(20,-1) \ArrowLine(-1,0)(1,0) 
\DashLine(-20,20)(-20,0){3} \DashLine(20,0)(20,20){3}   
\ArrowLine(20,0)(40,0) \Vertex(-20,0){3} \Vertex(20,0){3}  
\end{picture}   
   \hspace*{\fill} 
\begin{picture}(140,50)(-70,-20)    
\ArrowLine(-40,0)(-20,0) \DashLine(-20,20)(0,10){3} 
\DashLine(0,10)(20,0){3} \DashLine(-20,0)(0,10){3}  
\DashLine(0,10)(20,20){3} \Line(-20,1)(20,1) \Line(-20,-1)(20,-1) 
\ArrowLine(-1,0)(1,0) \ArrowLine(20,0)(40,0) 
\Vertex(-20,0){3} \Vertex(20,0){3}  
\end{picture}  
   \hspace*{\fill} 
\\ 
   \hspace*{\fill} 
\begin{picture}(10,10)(-5,-5)
\Text(0,5)[c]{(b)}   
\end{picture}
   \hspace*{\fill} 
\caption{\label{diagrams}%
Diagrams for  (a) $\,\Xi^0\pi^-\rightarrow\Xi^0\pi^-\,$   
and  (b) $\,\Xi^-\pi^0\rightarrow\Xi^-\pi^0.\,$   
The vertices are generated by  ${\cal L}^{\rm s}$  in  
Eq.~(\ref{L1strong}).  
A dashed line denotes a pion field, and a single (double) 
solid-line denotes a $\Xi$ ($\Xi^*$) field.}
\end{figure}
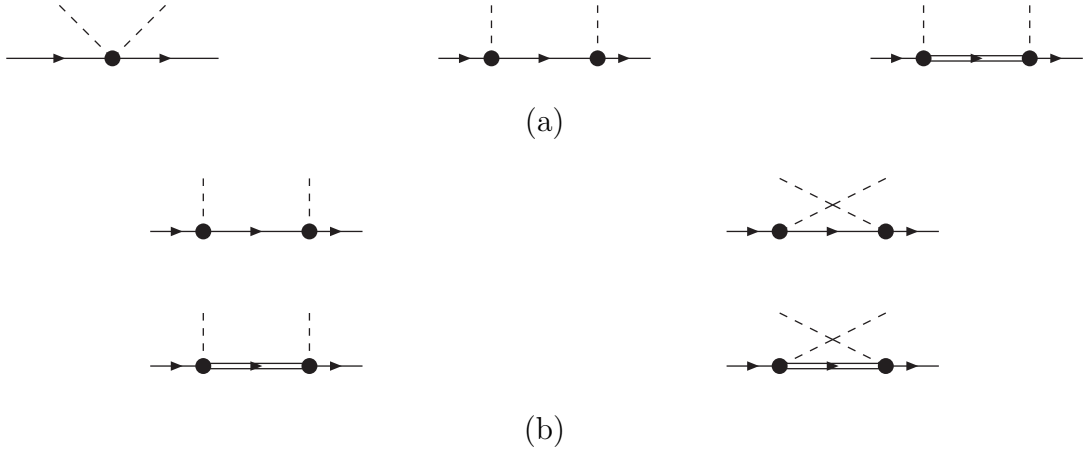

The scattering amplitudes for  $\,\Xi^0\pi^-\rightarrow\Xi^0\pi^-\,$   
and  $\,\Xi^-\pi^0\rightarrow\Xi^-\pi^0\,$  are derived from 
the diagrams shown in Figure~\ref{diagrams}.  
Of these, the first two diagrams in  Figure~\ref{diagrams}(a) 
and the first one in  Figure~\ref{diagrams}(b)  do not contribute 
to the  $\,J=3/2\,$  channel.  
From the rest of the diagrams, we can construct the amplitudes for the 
$\,I=1/2\,$  and $\,I=3/2\,$  channels, 
\begin{eqnarray}
\begin{array}{c}   \displaystyle   
{\cal M}_{I=1/2}^{}  \;=\;  
2{\cal M}_{\Xi^0\pi^-\rightarrow \Xi^0\pi^-}^{}  
\,-\,  {\cal M}_{\Xi^-\pi^0\rightarrow \Xi^-\pi^0}^{}   \;,  
\vspace{2ex} \\   \displaystyle
{\cal M}_{I=3/2}^{}  \;=\;  
-{\cal M}_{\Xi^0\pi^-\rightarrow \Xi^0\pi^-}^{}
\,+\,  2{\cal M}_{\Xi^-\pi^0\rightarrow \Xi^-\pi^0}^{}   \;,  
\end{array}
\end{eqnarray}
and project out the partial waves in the usual way.\footnote{%
See, e.g., Ref.~\cite{phases}.} 
Calculating the $\,J=3/2\,$  P-wave phases, and evaluating them 
at a center-of-mass energy equal to the $\Omega^-$ mass, yields 
\begin{eqnarray}   
\begin{array}{rcl}   \displaystyle   
\delta_1^{}  
&\!\approx&\!   \displaystyle   
{-|\bfm{k}|^3 m_\Xi^{}\over 24\pi f^2 m_\Omega^{}} \left[\,
{(D-F)^2\over m_\Omega^{}-m_\Xi^{}}  
\,+\,  
{\ratio{1}{2}\,{\cal C}^2\over m_\Omega^{}-m_{\Xi^*}^{}}_{}^{}  
\,+\,  
{\ratio{1}{18}\,{\cal C}^2\over m_\Omega^{}-2m_\Xi^{}+m_{\Xi^*}^{}}  
_{\vphantom{k}}^{\vphantom{k}}  
\,\right]   \;,  
\vspace{2ex} \\    \displaystyle       
\delta_3^{}  
&\!\approx&\!   \displaystyle   
{-|\bfm{k}|^3 m_\Xi^{}\over 24\pi f^2 m_\Omega^{}} \left[\, 
{-2(D-F)^2\over m_\Omega^{}-m_\Xi^{}}  
\,-\,  
{\ratio{1}{9}\,{\cal C}^2\over m_\Omega^{}-2m_\Xi^{}+m_{\Xi^*}^{}} 
_{\vphantom{k}}^{\vphantom{k}}  
\,\right]   \;.  
\end{array}   \label{strongph}  
\end{eqnarray}

\begin{figure}[ht]         
   \hspace*{\fill}   
\begin{picture}(300,220)(-50,-120)   
\LinAxis(0,90)(0,-90)(6,5,3,0,1) \LinAxis(200,90)(200,-90)(6,5,-3,0,1) 
\LinAxis(0,90)(200,90)(4,5,-3,0,1) \LinAxis(0,-90)(200,-90)(4,5,3,0,1)  
\DashLine(145.878,90)(145.878,-90){1} 
\SetWidth{1.0} 
\Curve{(0,0)(2.5,0.000361)(5.,0.0029)(7.5,0.00985)(10.,0.0236)
(12.5,0.0466)(15.,0.0817)(17.5,0.132)(20.,0.201)(22.5,0.293)
(25.,0.413)(27.5,0.566)(30.,0.76)(32.5,1.)(35.,1.3)
(37.5,1.68)(40.,2.14)(42.5,2.71)(45.,3.42)(47.5,4.31)
(50.,5.42)(52.5,6.83)(55.,8.63)(57.5,11.)(60.,14.1)
(62.5,18.5)(65.,24.8)(67.5,34.5)(70.,50.8)(72.5,82.5)
(72.8,89.3)} 
\Curve{(82.9,-89.7)(84.1,-79.2)(85.,-73.2)
(87.5,-61.5)(90.,-54.1)(92.5,-49.2)(95.,-45.7)(97.5,-43.1)
(100.,-41.2)(103.,-39.8)(105.,-38.7)(108.,-37.9)(110.,-37.3)
(113.,-36.8)(115.,-36.5)(118.,-36.3)(120.,-36.1)(123.,-36.1)
(125.,-36.1)(128.,-36.2)(130.,-36.3)(133.,-36.4)(135.,-36.6)
(138.,-36.9)(140.,-37.1)(143.,-37.4)(145.,-37.7)(148.,-38.1)
(150.,-38.4)(153.,-38.8)(155.,-39.2)(158.,-39.6)(160.,-40.)
(163.,-40.4)(165.,-40.8)(168.,-41.3)(170.,-41.7)(173.,-42.2)
(175.,-42.7)(178.,-43.1)(180.,-43.6)(183.,-44.1)(185.,-44.6)
(188.,-45.1)(190.,-45.6)(193.,-46.2)(195.,-46.7)(198.,-47.2)
(200.,-47.7)} 
\DashCurve{(0,0)(2.5,0.0000381)(5.,0.000304)(7.5,0.00102)(10.,0.00241)
(12.5,0.00469)(15.,0.00806)(17.5,0.0127)(20.,0.0188)(22.5,0.0265)
(25.,0.036)(27.5,0.0475)(30.,0.0609)(32.5,0.0765)(35.,0.0943)
(37.5,0.114)(40.,0.137)(42.5,0.162)(45.,0.189)(47.5,0.219)
(50.,0.252)(52.5,0.287)(55.,0.325)(57.5,0.365)(60.,0.408)
(62.5,0.453)(65.,0.501)(67.5,0.552)(70.,0.605)(72.5,0.661)
(75.,0.719)(77.5,0.78)(80.,0.843)(82.5,0.908)(85.,0.976)
(87.5,1.05)(90.,1.12)(92.5,1.19)(95.,1.27)(97.5,1.35)
(100.,1.43)(103.,1.52)(105.,1.6)(108.,1.69)(110.,1.78)
(113.,1.88)(115.,1.97)(118.,2.07)(120.,2.17)(123.,2.27)
(125.,2.37)(128.,2.47)(130.,2.58)(133.,2.69)(135.,2.8)
(138.,2.91)(140.,3.02)(143.,3.14)(145.,3.26)(148.,3.37)
(150.,3.49)(153.,3.62)(155.,3.74)(158.,3.86)(160.,3.99)
(163.,4.12)(165.,4.25)(168.,4.38)(170.,4.51)(173.,4.64)
(175.,4.78)(178.,4.91)(180.,5.05)(183.,5.19)(185.,5.33)
(188.,5.47)(190.,5.61)(193.,5.75)(195.,5.9)(198.,6.04)
(200.,6.19)}{4}  
\footnotesize  
\rText(-40,0)[][l]{$\delta_{2I}^{}\;(\rm degrees)$}   
\Text(100,-110)[t]{$|\bfm{k}|\;(\rm GeV)$}   
\Text(-5,90)[r]{$30$} \Text(-5,60)[r]{$20$} \Text(-5,30)[r]{$10$} 
\Text(-5,0)[r]{$0$} \Text(-5,-30)[r]{$-10$} \Text(-5,-60)[r]{$-20$} 
\Text(-5,-90)[r]{$-30$} 
\Text(0,-95)[t]{$0$} \Text(50,-95)[t]{$0.1$} \Text(100,-95)[t]{$0.2$} 
\Text(150,-95)[t]{$0.3$} \Text(200,-95)[t]{$0.4$} 
\end{picture} 
   \hspace*{\fill} 
\caption{\label{plot}%
Scattering phases as a function of the center-of-mass momentum of 
the pion. 
The solid and dashed curves denote  $\delta_1^{}$  and  $\delta_3^{}$, 
respectively.  
The vertical dotted-line marks the momentum in the
$\,\Omega^-\rightarrow\Xi\pi\,$  decay.}  
\end{figure}
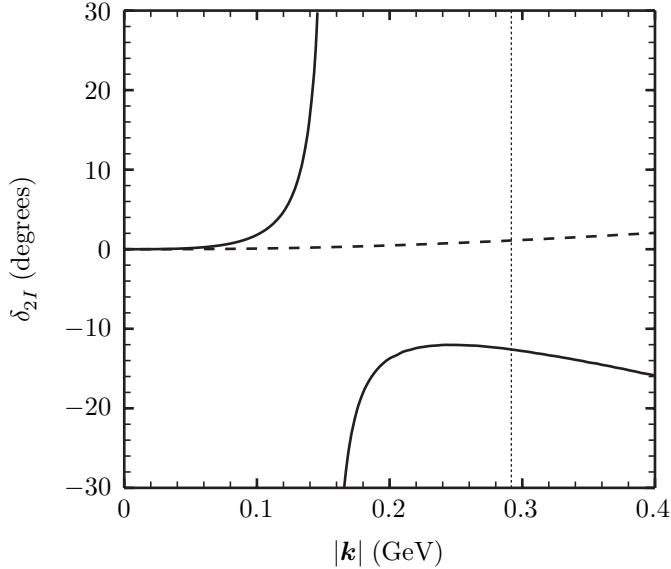

The phases are dominated by the terms proportional to ${\cal C}^2$  
arising from the  $\Xi^*\Xi\pi$  couplings. 
For this reason, we do not use the value  $\,{\cal C} \approx 1.5\,$  
obtained from a fit to decuplet decays at tree-level~\cite{manjen}, 
nor the value  $\,{\cal C}\approx 1.2\,$  obtained from a one-loop 
fit~\cite{butler}. 
Instead, we determine the value of  ${\cal C}$  from a tree-level fit 
to the width of the  $\,\Xi^*\ra\Xi\pi\,$  decay, which gives  
$\,{\cal C}=1.4 \pm 0.1.\,$ 
Using  $\,f=f_{\!\pi}^{}\approx 92.4\;\rm MeV,\,$  
isospin-symmetric masses, and the values  $\,D=0.61\,$  and  
$\,F=0.40,\,$  we obtain\footnote{%
We have also computed the phases in chiral perturbation 
theory without treating the baryons as heavy, and found very similar 
results,  $\,\delta_1^{}=-13.1^{\small\rm o}\,$  and    
$\,\delta_3^{}=1.4^{\small\rm o}.\,$}  
\begin{eqnarray}   
\delta_1^{}  \;=\;  -12.8^{\small\rm o}  
\;, \hspace{3em}   
\delta_3^{}  \;=\;  1.1^{\small\rm o}   \;.    
\end{eqnarray}  
In Figure~\ref{plot}  we plot the scattering phases as a function 
of the pion momentum. 

Our estimate indicates that the  $\,I=1/2\,$  P-wave phase for the  
$\Xi\pi$  scattering is larger than other baryon-pion 
scattering phases.   
Eq.~(\ref{strongph}) shows that this phase is dominated by 
the $s$-channel  $\Xi^*$-exchange diagram.  
This is what one would expect from the fact that the $\Xi^*$ shares 
the quantum numbers of the channel. 
Notice, however, that the  phase is not large due to the resonance 
because it is evaluated at a center-of-mass energy equal to  
the  $\Omega^-$ mass,  significantly above the  $\Xi^*$  pole.  
The phase is relatively large because the pion momentum in  $\Omega^-$ 
decays is large.\footnote{In fact, this  $\Xi\pi$-scattering phase  
is much larger than the corresponding  P-wave  $\Lambda\pi$-scattering 
phase  $\,\delta_P\approx -1.7^{\small\rm o}\,$~\cite{wisa}  because the pion 
momentum is
much larger  in the reaction  $\,\Omega^-\rightarrow\Xi\pi\,$  than it is in 
the reaction  $\,\Xi\rightarrow\Lambda\pi$.}

\section{Estimate of the weak phases}

Within the standard model the weak phases  $\phi_{1}$  and  $\phi_{3}$  
arise from the $C\!P$-violating phase in the CKM matrix. 
The short-distance effective Hamiltonian describing 
the $\,|\Delta S| =1\,$  weak interactions in the 
standard model can be written as
\begin{equation}
{\cal H}_{\rm eff}  \;=\;   
{G_{\rm F}^{}\over\sqrt{2}}\, V^*_{ud} V_{us}^{} 
\sum_i C_i^{}(\mu)\, Q_i^{}(\mu)   
\;+\;  {\rm h.c.}   \;,   
\end{equation}   
where the sum is over all the  $Q_i^{}(\mu)$  four-quark operators, 
and the  $\,C_i^{}(\mu)=z_i^{}(\mu)+\tau y_i^{}(\mu)\,$  are   
the Wilson coefficients, with  
$\,\tau=-V^*_{td} V_{ts}^{}/V^*_{ud} V_{us}^{}.\,$  
We use the same operator basis of Ref.~\cite{steger} because our 
calculation will parallel that one, but we use the latest values for 
the Wilson coefficients from Ref.~\cite{buras}. 
To calculate the phases, we write 
\begin{equation}
\ri{\cal M}_{\Omega^-\rightarrow\Xi\pi}^{}  \;=\;
-\ri{G_{\rm F}^{} \over \sqrt{2}} \, V^*_{ud} V_{us}^{} 
\sum_i C_i^{}(\mu)\, \langle \Xi\pi| Q_i^{}(\mu) |\Omega^- \rangle   \;.
\end{equation}
Unfortunately, we cannot compute the matrix elements of the 
four-quark operators in a reliable way. 
As a benchmark, we employ the vacuum-saturation method used 
in Ref.~\cite{steger}. 
For  $\,\Omega^-\rightarrow\Xi^0 \pi^-,\,$  we obtain
\begin{equation} 
{\cal M}_{\Omega^-\rightarrow\Xi^0 \pi^-}^{}  \;=\;
-{G_{\rm F}^{}\over\sqrt{2}}\, V_{ud}^* V_{us}^{} 
\left( M_1^{\rm P}+M_3^{\rm P} \right) 
\langle \Xi^0 \bigl| \bar{u}\gamma^\mu\gamma_5^{} s 
\bigr| \Omega^- \rangle
\langle \pi^- \bigl| \bar{d}\gamma_\mu^{}\gamma_5^{} u 
\bigr| 0 \rangle   \;,  
\end{equation}  
where we have used the notation 
\begin{eqnarray}   
M^{\rm P}_1  &\!\!=&\!\!      
\ratio{1}{3} \bigl( C_1^{} - 2 C_2^{} \bigr) - \ratio{1}{2}\, C_7^{} 
\,+\, 
\xi \left[\, \ratio{1}{3} \bigl( -2 C_1^{} + C_2^{} \bigr) - C_3^{} 
           - \ratio{1}{2}\, C_8^{} \,\right]   
\nonumber \\ && \!\!\! \!   
+\; 
{2 m^2_\pi\over (m_u^{}+m_d^{}) (m_u^{}+m_s^{})} 
\left[ C_6^{} + \ratio{1}{2}\, C_8^{} 
       + \xi \left( C_5^{} + \ratio{1}{2}\, C_7^{} \right) \right]   \;, 
\\   
\nonumber \\
M^{\rm P}_3 &\!\!=&\!\!   
-\ratio{1}{3} (1+\xi) (C_1^{}+C_2^{}) \,+\,  \ratio{1}{2}\, C_7^{}  
\,+\,  \ratio{1}{2}\, \xi\, C_8^{} 
\,+\,  {m^2_\pi\over(m_u^{}+m_d^{})(m_u^{}+m_s^{})}  
      \left( \xi C_7^{} + C_8^{} \right)   \;.   
\end{eqnarray}   
The current matrix-elements that we need are found from 
the leading-order strong Lagrangian in  Eq.~(\ref{L1strong})  to be 
\begin{eqnarray} 
\langle \Xi^0 \bigl| \bar{u}\gamma^\mu\gamma_5^{} s 
\bigr| \Omega^- \rangle
\;=\;  -{\cal C}\, \bar{u}_\Xi^{}\, u_\Omega^\mu   
\;, \hspace{3em}   
\langle \pi^- \bigl| \bar{d}\gamma_\mu^{}\gamma_5^{} u 
\bigr| 0 \rangle
\;=\;  \ri\sqrt{2}\,f_{\!\pi}^{} k_\mu^{}   \;,  
\end{eqnarray}  
and from these we obtain the matrix elements for pseudoscalar 
densities as  
\begin{eqnarray}
\begin{array}{c}   \displaystyle      
\langle \Xi^0 \bigl| \bar{u}\gamma_5^{} s \bigr| \Omega^- \rangle  
\;=\;  
{{\cal C}\over m_u^{}+m_s^{}}\, \bar{u}_\Xi^{}\,k_\mu^{}\, u_\Omega^\mu 
\vspace{2ex} \\    \displaystyle       
\langle \pi^- \bigl| \bar{d}\gamma_5^{} u \bigr| 0 \rangle
\;=\;  \ri\sqrt{2}\, f_{\!\pi}^{}\, {m^2_\pi \over m_u^{}+m_d^{}}   \;.   
\end{array}     
\end{eqnarray}     
Numerically, we will employ  
$\,m_\pi^2 / \left[ (m_u^{}+m_d^{})(m_u^{}+m_s^{}) \right] \sim 10$,  
$\,\xi=1/N_{\rm c}^{}=1/3,\,$  and the Wilson coefficients from 
Ref.~\cite{buras}
that correspond to the values  $\,\mu=1$~GeV,  $\,\Lambda=215$~MeV,  
and  $\,m_t=170$~GeV. 
Given the crudeness of the vacuum-insertion method, 
we use the leading-order Wilson coefficients. 
For the CKM angles, we use the Wolfenstein parameterization 
and the numbers  $\,\lambda =0.22$,  $\,A=0.82$, 
$\,\rho=0.16\,$  and  $\,\eta=0.38\,$~\cite{mele}.  
Putting all this together, we find
\begin{eqnarray}
\begin{array}{c}   \displaystyle      
\alpha_{3}^{(\Omega)} \re^{{\rm i}\phi_3^{}}  \;=\;  
-0.11  \,+\,  2.8\times10^{-6}\, \ri   \;,
\vspace{2ex} \\    \displaystyle       
\alpha_{1}^{(\Omega)} \re^{{\rm i}\phi_1^{}}  \;=\;  
0.23  \,+\,  2.3\times 10^{-4}\, \ri   \;.
\end{array}
\end{eqnarray}
The  $\,|\Delta\bfm{I}|=3/2\,$  amplitude predicted in vacuum 
saturation is comparable to the one we extract from the data,
$\,\alpha_3^{(\Omega)} = -0.07\pm 0.01.\,$  
To estimate the weak phase, we can obtain the real part of the 
amplitude from experiment and the imaginary part of the amplitude 
from the vacuum-saturation estimate to get 
$\,\phi_3^{}\approx -4\times 10^{-5}.\,$  
Unlike its  $\,|\Delta\bfm{I}|=3/2\,$  counterpart, 
the $\,|\Delta\bfm{I}|=1/2\,$  amplitude is predicted to be about 
a factor of four below the fit.\footnote{%
We note here that only the relative sign between  $\alpha_{1}^{(\Omega)}$  
and  $\alpha_{3}^{(\Omega)}$  is determined, while the overall sign 
of either the predicted or experimental numbers is not.}  
Taking the same approach as that in estimating  $\phi_3^{}$  results in 
$\,\phi_1^{}\approx 3\times 10^{-4}.\,$  
We can also take the phase directly from the vacuum-saturation estimate 
(assuming that both the real and imaginary parts of the amplitude 
are enhanced in the same way by the physics that is missing from this 
estimate) to find  $\,\phi_1^{} = 0.001.\,$

For the decay of the  $\Omega^-$,  it is much more difficult 
to estimate the phases in quark models than it is for other 
hyperon decays. 
For instance, to calculate the phase of the  $\,|\Delta\bfm{I}|=1/2\,$   
amplitude, we would need to calculate the matrix element  
$\,\langle \Xi^{*-}|H_W|\Omega^- \rangle,\,$   
but this vanishes for the leading  $\,|\Delta\bfm{I}|=1/2\,$   operator 
because the quark-model wavefunctions of the  $\Omega^-$  and  
the  $\Xi^{*-}$  do not contain $u$-quarks. 
Considering only valence quarks, these models would then predict that 
the phase is equal to the phase of the leading penguin operator,\footnote{%
Early calculations obtain the amplitude as a sum of a bag model estimate 
of the penguin matrix element and factorization 
contributions~\cite{quarkmodels}.} or about  $\,\phi_1^{}\sim 0.006$.

\section{Results and Conclusion}

Finally, we can collect all our results to estimate the $C\!P$-violating 
rate asymmetry  $\Delta \bigl( \Xi^0\pi^- \bigr) $.  
They are      
\begin{eqnarray}
\begin{array}{rcl}   \displaystyle   
{\alpha_{3}^{(\Omega)}\over \alpha_{1}^{(\Omega)}}  &\!\approx&\!  
-0.07   \;,  
\vspace{2ex} \\    \displaystyle       
|\sin \bigl( \delta_3^{}-\delta_1^{} \bigr) |  &\!\approx&\!  0.24   \;,  
\vspace{2ex} \\    \displaystyle       
|\sin(\phi_3^{}-\phi_1)|  &\!\approx&\!  
3\times 10^{-4} \;~{\rm or}~\; 0.001   \;,  
\end{array}   \label{numbers}
\end{eqnarray}  
where the first number for the weak phases corresponds to the 
conservative approach of taking only the imaginary part of the 
amplitudes from the vacuum-saturation estimate and the second 
number is the phase predicted by the model.  
The difference between the resulting numbers, 
$\, \bigl| \Delta \bigl( \Xi^0\pi^- \bigr) \bigr| = 
   7\times 10^{-6} \;~{\rm or}~\; 2\times 10^{-5} ,\,$  
can be taken as a crude measure of the uncertainty in the evaluation of 
the weak phases.  
For comparison, estimates of rate asymmetries in the octet-hyperon 
decays~\cite{donpa} result in values of less than~$\,10^{-6}.$

A model-independent study of $C\!P$ violation beyond the standard 
model in hyperon decays was done in  Ref.~\cite{heval}.  
We can use those results to find that the $C\!P$-violating rate asymmetry 
in  $\,\Omega^-\rightarrow\Xi^0\pi^-\,$  could be ten times larger than  
our estimate above if new physics is responsible for $C\!P$ violation.
The upper bound in this case arises from the constraint imposed  on 
new physics by the value of  $\epsilon$  because the P-waves involved 
are parity conserving.
  
In conclusion, we find that the $C\!P$-violating rate 
asymmetry in  $\,\Omega^-\rightarrow\Xi^0\pi^-\,$  is about   
$\,2 \times 10^{-5}\,$  within the standard model.  
Although there are significant uncertainties in our estimates,  
it is probably safe to say that the rate asymmetry in  
$\,\Omega^-\rightarrow\Xi\pi\,$  decays is significantly larger 
than the corresponding asymmetries in other hyperon decays.

\newpage 

\noindent {\bf Acknowledgments} This work  was supported in
part by DOE under contract number DE-FG02-92ER40730. 
We thank the Department of Energy's Institute for Nuclear Theory 
at the University of Washington for its hospitality and for 
partial support. We also thank John F. Donoghue for helpful discussions.

\end{document}